\documentclass[prd,twocolumn,aps,superscriptaddress,showpacs]{revtex4}
\usepackage{amssymb}
\usepackage{amsmath,bm}
\usepackage{graphicx}
\usepackage[normalem]{ulem}
\usepackage[dvips]{color}
\usepackage{subfigure}

\renewcommand\sout{\bgroup \color{red} \ULdepth=-.5ex \ULset}

\begin{document}

\title{Isovector properties of quark matter and quark stars in an isospin-dependent confining model}
\author{Peng-Cheng Chu}
\affiliation{Science school, Qingdao University of technology, Qingdao 266000, China }
\author{Lie-Wen Chen\footnote{%
Corresponding author: lwchen$@$sjtu.edu.cn}}
\affiliation{School of Physics and Astronomy and Shanghai Key Laboratory for
Particle Physics and Cosmology, Shanghai Jiao Tong University, Shanghai 200240, China }
\affiliation{Center of Theoretical Nuclear Physics, National Laboratory of Heavy-Ion
Accelerator, Lanzhou, 730000, China}
\date{\today}

\begin{abstract}
The confining quark matter (CQM) model, in which the confinement and asymptotic freedom are modeled
via the Richardson potential for quark-quark vector interaction and the chiral symmetry restoration at high density
is described by the density dependent quark mass, is extended to include isospin dependence of the
quark mass. Within this extended isospin-dependent confining quark matter
(ICQM) model, we study the properties of strange quark matter and quark stars. We find
that including isospin dependence of the quark mass can significantly
influence the quark matter symmetry energy, the stability of strange quark matter and
the mass-radius relation of quark stars. In particular, we demonstrate although
the recently discovered large mass pulsars PSR J1614.2230 and PSR J0348+0432 with
masses around two times solar mass ($2M_{\odot}$) cannot be quark stars within the original CQM model,
they can be well described by quark stars in the ICQM model if the isospin dependence of quark mass
is strong enough so that the quark matter symmetry energy is about four times that
of a free quark gas. We also discuss the effects of the density dependence of quark mass on
the properties of quark stars.
Our results indicate that the heavy quark stars with mass around $2M_{\odot}$ (if exist)
can put strong constraints on isospin and density dependence of the quark mass
as well as the quark matter symmetry energy.
\end{abstract}

\maketitle

\section{Introduction}

The investigation of the properties of {strongly interacting} matter is one of the fundamental issues in the current research frontiers of nuclear physics, astrophysics, and cosmology. In particular, the equation of state (EOS) of {strongly interacting} matter plays a central role in understanding the nuclear structures and reactions, many important issues in astrophysics, and the matter state at early universe.
It has been established that quantum chromodynamics (QCD) is the fundamental theory of strong interaction and it provides the basis for understanding the properties of {strongly interacting} matter.
Perturbative QCD (pQCD) is successful in describing the processes or systems with large energy scales, such as high energy collision processes and the systems with extremely large baryon densities, while it still has difficulty in treating
the physics in low energy nuclear physics or system with moderate densities such as neutron stars due to
the complicated nonperturbative QCD features~\citep{Fukushima11}.
On the other hand, the ab initio numerical Monte Carlo simulations of lattice QCD (LQCD) provide a straight way to obtain the properties of {strongly interacting} at finite temperature with zero baryon density (baryon chemical potential), but the LQCD still cannot be applied in the case of finite baryon chemical potential because of the famous sign problem~\citep{Barbour86}. Therefore, it is still a big challenge to evaluate theoretically the properties of {strongly interacting} matter at finite baryon density, especially at supra-saturation densities.

Experimentally, heavy-ion collisions provide a unique approach to investigate the properties of {strongly interacting} matter in terrestrial laboratories. The heavy-ion collision experiments at Relativistic Heavy Ion Collider (RHIC) and Large Hadron Collider (LHC), for instance, have revealed many interesting features of {strongly interacting} matter at zero baryon density and high temperatures.
The properties of {strongly interacting} matter at high baryon densities and finite temperatures can be explored by
the beam-energy scan program at RHIC as well as future experiments planned in the Facility for Antiproton and Ion Research (FAIR)
at GSI and the Nuclotron-based Ion Collider Facility (NICA) at JINR.
On the other hand, compact stars provide a natural site to investigate the properties of {strongly interacting} matter at high baryon densities and low temperatures (or finite temperatures like proto-compact stars). Neutron stars (NSs) have been shown to provide the testing grounds for EOS of neutron-rich nuclear matter~\citep{Lattimer04,Steiner05}. Theoretical studies~\citep{Bombaci04,Staff07,Herzog11} suggest that NSs may be converted to strange quark stars (QSs), which are entirely made up of deconfined absolutely stable u, d and s quark matter with some leptons, i.e., strange quark matter (SQM). An important feature of QSs is that they usually have a smaller radii than NSs~\citep{Kapoor01} for a fixed star mass, especially for the small mass stars. The possible existence of QSs has important implications for understanding the properties of SQM which determine the structure of QSs~\citep{Ivanenko69,Itoh70,Bodmer71,Witten84,Farhi84,Alcock86,Weber05}.
Recently, the masses of two massive pulsars PSR J1614-2230~\citep{Demorest10} and PSR J0348+0432~\citep{Antoniadis13} have been precisely {determined to be around 2$M_{\odot}$}, which provide us the new record for the maximum mass of pulsars. In order to describe these two stars as QSs, the interaction between quarks should be very strong~\cite{Alford03,Baldo03,Ruster04,Alford05,Ippolito08,Lai11,Avellar11,Bonanno12}, which is remarkably consistent with the finding that quarks and gluons form a strongly interacting system in high energy heavy ion collisions.

Since we cannot determine the properties for SQM in compact stars with large baryon chemical potential by using ab initio pQCD and LQCD, many phenomenological QCD-inspired effective quark matter models have been proposed to calculate the properties of quark star matter, such as the MIT bag model ~\cite{Cho74,Far84,Alc86,Alf05,Weber05}, the Nambu-Jona-Lasinio (NJL) model~\citep{Rehberg96,Han01,Rus04,Men06}, the pQCD approach~\cite{Fre77,Fra01,Fra05,Kur10}, the Dyson-Schwinger approach~\cite{Rob94,Zon05,Qin11}, the CDDM model ~\cite{Fow81,Cha89,Cha91,Cha93,Cha96,Ben95,Pen99,Pen00,Pen08,Li11,
Zhang02,Zhang03, Wen05, Mao06,Wu08,Yin08}, and the quasi-particle model~\citep{Sch97,Sch98,Pes00,Hor04,Alford07}. {For the phenomenological models of quark matter, how to deal with the quark confinement is a basic problem}. The MIT bag model and its density dependent versions provide a popular way to treat quark confinement. The other popular way to treat quark confinement is to vary the interaction part of quark mass, such as the CDDM model and the quasi-particle model with density dependent equivalent quark mass~\citep{Chu14,Fow81,Cha89,Cha91,Cha93,Cha96,Ben95,Pen99,Pen00,Pen08,Zhang02,Zhang03,Wen05,Mao06,Wu08,Yin08}.
In addition, another possible way of dealing with quark confinement is to consider the interquark potential which originates from gluon exchanges. Since 't Hooft~\citep{tHooft74} suggested to use the inverse of the color number as an expansion parameter, large baryon number density system like compact stars can be calculated in tree level with a meson sector phenomenological interquark potential~\citep{Witten79}. In this aspect, the confining quark matter (CQM) model~\citep{Dey98,Dey991,Dey992,Dey993} provides a good choice to describe the quark star matter. In this model, the quarks interact through the Richardson potential~\citep{Ric79,Sinha13} where the asymptotic freedom and confinement are considered, and the chiral symmetry feature of QCD is described by the density dependent quark mass.

During the last decades, isospin physics has attracted much interest, mainly due to the fact that rare isotopes with extreme isospin can be produced and used to induce heavy-ion collisions in terrestrial laboratories and also large isospin can be appeared in neutron stars (see, e.g., Ref.~\cite{Li14}, for a review). Since the u-d quark isospin asymmetry could be large in QSs and the u-d quark numbers are generally unequal in high energy HICs at RHIC/LHC/FAIR/NICA, the isovector properties of quark matter may play an important role in these issues.
In recent years, great efforts have been devoted to exploring the QCD phase diagram of {strongly interacting} matter at finite isospin~\citep{Son01,Frank03,Toublan03,Kogut04,He05,Hatsuda94,Di06,Zhang07,Bielich10,Shao12}, and this is very useful for understanding the properties of QSs, QCD phase diagram at extreme isospin condition, and isospin effects in high energy HICs. For asymmetric nucleonic matter, the symmetry energy plays a critical role in many issues of nuclear physics and astrophysics (see, e.g., Ref.~\citep{Li14,Chen17}). For isospin asymmetric quark matter, the isovector properties can significantly affects the QCD phase boundary and the particle fractions in compact stars, see, e.g., the very recent work based on the (P)NJL model~\citep{Liu16}.

One important issue about isospin asymmetric quark matter is the isospin splitting of the in-medium quark mass, which is directly related to isospin dependent chiral condensates. The chiral condensate is an order parameter of spontaneous chiral symmetry
breaking in QCD. The isospin dependence of chiral condensates in asymmetric nucleonic matter has been extensively investigated in isospin nuclear physics (see, e.g., Ref.~\citep{Cai17} and references therein) and it has been shown that the chiral condensates are isospin dependent, suggesting the isospin dependence of quark mass in asymmetric quark matter. The original CQM model does not consider the isospin dependence of quark mass, and it is thus interesting to see the effects of isospin dependence of the quark mass in the CQM model, which is the main motivation of the present work.

In this work, we extend the CQM model to include isospin dependence of the quark mass. Based on the isospin dependent model, we will investigate the stability of SQM, the mass-radius relation for QSs, the quark matter symmetry energy and the isospin splitting of quark mass in asymmetric quark matter.

\section{The theoretical model}
In the original CQM model~\citep{Dey98}, the model Hamiltonian of describing u-d-s quark matter at zero temperature and finite chemical potential is given by
\begin{eqnarray}
\mathcal{H}=\sum\limits_i(\alpha_i\cdot p_i+\beta_iM_i)+\sum\limits_{i<j}\frac{\lambda(i)\lambda(j)}{4}V_{ij},
\end{eqnarray}
where $i$ and $j$ stand for quark flavors, $\alpha_i$ and $\beta_i$ come from Dirac equation, $\lambda_i$ is SU(3) matrix for interacting quarks, $V_{ij}$ is the quark vector interaction and taken as the Richardson potential, and $M_i$ is the quark mass which is density dependent and parameterized as
\begin{eqnarray}
M_i=m_i+(310\text{MeV})\text{sech}(\nu\frac{n_B}{n_0}),
\label{mCQM}
\end{eqnarray}
where $i$ stands for the flavor of the quarks, $m_i$ represents the current quark mass ($m_u=4$ MeV, $m_d=7$ MeV and $m_s=150$ MeV in the original CQM model), $n_B$ is the baryon number density, $n_0=0.17$ fm$^{-3}$ is nuclear matter normal (saturation) density, and $\nu$ is a parameter determining the density dependence for quark mass.

One can see that the quark mass decreases smoothly from $310$ MeV to the current mass when the baryon density varies from zero to a high value.
Eq.~(\ref{mCQM}) also indicates that the value of $M_i$ for u quark is exactly the same as that of d quark, suggesting that there is no isospin dependence in this mass term and the quark-quark interaction is identical for different flavors of quarks. Since the quark mass should be isospin dependent {in asymmetric quark matter}, one can then consider isospin dependence in the quark mass in Eq.~(\ref{mCQM}) and extend the CQM model to isospin dependent confining quark matter (ICQM) model. It should be mentioned that in relativistic mean field (RMF) model, a similar mass term as in Eq.~(\ref{mCQM}) appears due to the isoscalar-scalar $\sigma$ meson field, and this mass $M_i$ is usually denoted as the Dirac mass which is calculated through the nucleon scalar self-energy in the Dirac equation for nucleons~\citep{Chen07}. For quark matter models including NJL-type Lagrangians, we can also obtain the quark mass term $M_i$ from quark self-energy~\citep{buballa05,Huang02}. In SU(2) NJL model, the quark condensate is given by $\langle \bar{q}q\rangle=-\frac{M_q-m_q}{2G}$~\citep{buballa05,Huang02}, where $M_q$ is the quark mass, $m_q$ is the current mass, and $G$ is the coupling constant for isoscalar-scalar field ($\sigma$ field). In SU(3) NJL model, the quark condensate cannot be easily obtained by this equation~\citep{buballa05,Chu15,Chu15b}, because all the condensates for different flavors of quarks are all {involved} and mixed in the term $(M_q-m_q)$. When the isovcetor-scalar field is considered in the RMF model~\citep{LiuB02,Chen07} or NJL model~\citep{Liu16,buballa05,Chu15,Chu15b}, the nucleon or quark Dirac mass then becomes isospin dependent.

In order to extend the CQM model to include isospin dependence of the quark mass, one can include the contribution of isovector-scalar channels into $M_i$ in Eq.~(\ref{mCQM}). Since the form of isospin dependence of the quark mass is unclear, we adopt in this work the phenomenological parameterization for isospin dependence of the equivalent quark mass in the CIDDM model~\citep{Chu14}, and then the quark mass is expressed as
\begin{eqnarray}
M_{i}&=&m_{i}+m_i^*\text{sech}(\nu_i\frac{n_B}{n_0})-\tau_i\delta D_I n^{\alpha}_B e^{-\beta n_B},
\label{mqiso}
\end{eqnarray}
where $D_I$, $\alpha>0$ and $\beta>0$ are parameters introducing isospin dependence of the quark mass in quark matter,
$\tau_i$ is the isospin quantum number for quarks and we set $\tau_i = 1$ for $i=u$ ($u$ quarks),
$\tau_i = -1$ for $i=d$ ($d$ quarks), and $\tau_i = 0$ for $i=s$ ($s$ quarks).
The isospin asymmetry $\delta$ is defined as
\begin{eqnarray}
\delta=3\frac{n_d-n_u}{n_d+n_u}
\end{eqnarray}
which has been extensively used in the literature~\citep{Chu14,Di06,Bielich10,Di10,Chen17,Shao12}.
One can see $\delta = 1$ ($-1$) for quark matter converted by pure neutron (proton) matter, consistent with the definition of isospin asymmetry for nuclear matter.
For two-flavor $u$-$d$ quark matter, one sees from Eq.~(\ref{mqiso}) that, the chiral symmetry is restored at high
density due to $\lim_{n_B\to\infty}M_{i=u,d} = 0$, if the current masses of $u$ and $d$
quarks are neglected. In addition, the quark mass in Eq.~(\ref{mqiso})
also satisfies the exchange symmetry between $u$ and $d$ quarks, which is required
by isospin symmetry of the strong interaction. Therefore, the phenomenological
parametrization form of the isospin dependent quark mass in
Eq.~(\ref{mqiso}) is quite general and respects the basic features of QCD.

In the CQM model~\citep{Dey98}, the quark vector interaction $V_{i,j}$ originates from gluon exchanges and it is taken as the Richardson potential~\citep{Ric79}, i.e.,
\begin{eqnarray}
V_{ij}&=&\frac{4\pi}{9}\frac{1}{\ln{(1+[( \mathbf{k}_i- \mathbf{k}_j)^2+D^{-2}]/\Lambda^2)}}\notag\\
&\times&\frac{1}{(\mathbf{k}_i- \mathbf{k}_j)^2+D^{-2}},
\end{eqnarray}
where $\mathbf{k}_i- \mathbf{k}_j$ means the momentum transfer between the $i$-th and $j$-th particles, $D$ is the screening length, and $\Lambda$ is the scale parameter. This potential will be screened in the medium due to pair creation and infrared divergence, and to the lowest order the squared inverse screening length (the gluon mass) can be expressed as~\citep{Kap79}
\begin{eqnarray}
(D^{-1})^2=\frac{2\alpha_0}{\pi}\sum\limits_{i=u,d,s}k_i^f\sqrt{(k_i^f)^2+M_i^2},
\label{Dscr}
\end{eqnarray}
where $k_i^f=(\pi^2 n_i)^{1/3}$ is the quark Fermi momentum, with $n_i$ the quark number density, and $\alpha_0$ is the perturbative quark gluon coupling constant. One can see that $D^{-1}$ satisfies the $u$-$d$ quark isospin exchange symmetry, and the potential incorporates asymptotic freedom and quark confinement, which are the basic features of QCD. In the present work, instead of using the average inverse screening length as in the original CQM model~\citep{Dey98} to simplify the numerical calculation, we sum over all the flavors of quarks to calculate the inverse screening length in Eq.~(\ref{Dscr}). We also adopt the original value of the scale parameter $\Lambda=100$ MeV and $\alpha_0 = 0.2$, which are obtained from pQCD calculations for hadron phenomenology.
One can find that the Richardson potential $V_{ij}$ becomes to infinity when the number density of quarks is zero, which indicates the confinement for quark interaction in QCD, while $V_{ij}$ decreases with density and restores asymptotic freedom (deconfinement) at high density.

In the present work, for the current mass $m_{i}$ of quarks in Eq.~(\ref{mqiso}), we set $m_u=m_d=5.5$ MeV and $m_s=95$ MeV.
In Eq.~(\ref{mqiso}), we also set $m_u^*=m_d^*=329.5$ MeV and $m_s^*=432 $ MeV in order to match the vacuum values of quark masses (constituent quark masses) of $M_{u0}=M_{d0}=335$ MeV and $M_{s0}=527$ MeV obtained in SU(3) NJL model with the popular parameter set HK~\citep{Hatsuda94}. In addition, the values of $\alpha>0$ and $\beta>0$ are fixed to be $\alpha=1.5$ and $\beta=1~\text{fm}^3$ throughout this paper to obtain a reasonable baryon density dependence for quark matter symmetry energy, which will be further discussed in Sec.~\ref{Results}.
In this work, for simplicity, we mainly focus on the case of $\nu_{u} =\nu_{d} \equiv \nu_{ud}$ and $\nu_{ud} = \nu_s = \nu$, and thus we have only two free parameters, namely, $\nu$ and $D_I$. We will give a brief discussion about the more general case with unequal $\nu_{ud}$ and $\nu_s$.

\section{Properties of Quark Matter}

For asymmetric quark matter, the EOS of quark matter consisting of $u$, $d$ and $s$ quarks, which is defined by its binding energy per baryon number, can be expanded in isospin asymmetry $\delta$ as
\begin{equation}
E(n_B ,\delta)=E_{0}(n_B)+E_{\mathrm{sym}}(n_B)\delta ^{2}+\mathcal{O}(\delta ^{4}),
\label{esym}
\end{equation}
where $E_{0}(n_B)=E(n_B ,\delta =0)$ is the binding energy per baryon number with an equal fraction of $u$ and $d$ quarks, and the quark matter symmetry energy is expressed as
\begin{eqnarray}
E_{\mathrm{sym}}(n_B) &=&\left. \frac{1}{2!}\frac{\partial ^{2}[(\epsilon_k+\epsilon_v)/n_B]}{\partial \delta ^{2}}\right\vert _{\delta =0},
\label{QMEsym}
\end{eqnarray}%
where $\epsilon_k$ and $\epsilon_v$ represent, respectively, the kinetic and potential parts of the energy density of the quark matter. One can see from Eq.~(\ref{esym}) the definition of quark matter symmetry energy is similar with that of nuclear matter symmetry energy~\cite{Li08,Chen11}.

For $u$-$d$-$s$ quark matter, the kinetic part of the energy density can be written as
\begin{eqnarray}
\epsilon_k&=&\frac{6}{(2\pi)^3}\sum\limits_{i=u,d,s}\int_0^{k_i^f} \text{d}^3k\sqrt{k^2+M_i^2}\notag\\
&=&\frac{3}{4\pi^2}\sum\limits_{i=u,d,s}\Big[k_i^f((k_i^f)^2+M_i^2/2)\sqrt{(k_i^f)^2+M_i^2}\notag\\
&-&\frac{M_i^4}{2}\ln{\frac{\sqrt{(k_i^f)^2+M_i^2}+k_i^f}{M_i}}\Big]\notag\\
&=&\frac{3}{4}\sum\limits_{i=u,d,s}\Big[n_i\sqrt{{k_i^f}^2+M_i^2}+M_i\rho_s^i\Big],
\end{eqnarray}
where $\rho_s^i$ is the quark scalar density given by
\begin{eqnarray}
\rho_s^i=&\frac{3}{2\pi^2}\Big[M_ik_i^f\sqrt{(k_i^f)^2+M_i^2}&\notag\\ &-M_i^3\ln{\frac{\sqrt{(k_i^f)^2+M_i^2}+k_i^f}{M_i}}&\Big].
\end{eqnarray}
The potential part of the energy density for $u$-$d$-$s$ quark matter can be obtained as
\begin{eqnarray}
\epsilon_v=&-&\frac{1}{2\pi^3}\sum\limits_{i,j}\int_{-1}^1\text{d}x \int_0^{k_j^f} k_j^2\int_{0}^{k_i^f} k_i^2\notag\\
&\times& f(k_i,k_j,M_i,M_j,x) V_{ij}\text{d}k_j\text{d}k_i,
\end{eqnarray}
where $f$ is defined as
\begin{eqnarray}
f(k_i,&k_j&,M_i,M_j,x)=\left( e_i\cdot e_j+2\cdot k_i\cdot k_j \cdot x +\frac{k_i^2 k_j^2}{e_i \cdot e_j}\right)\notag\\
 &\times & \frac{1}{(e_i-M_i)(e_j-M_j)}
\end{eqnarray}
with
\begin{eqnarray}
e_i=\sqrt{k_i^2+M_i^2}+M_i.
\end{eqnarray}

The chemical potential for each flavor of quarks can then be obtained as
\begin{eqnarray}
\mu_i= \mu_{i,K} +\mu_{i,V},
\end{eqnarray}
where $\mu_{i,K}$ and $\mu_{i,V}$ are the contributions from the kinetic and potential parts of the energy density, respectively.
The $\mu_{i,K}$ can be expressed as
\begin{eqnarray}
 \mu_{i,K} =\frac{\partial \epsilon_k}{\partial M_i}\frac{\partial M_i}{\partial n_i}=\sqrt{M_i^2+(k_i^f)^2}+(\rho_s^u+\rho_s^d+\rho_s^s)\frac{\partial M_i}{\partial n_i},
\end{eqnarray}
and the contribution from the potential part of the energy density can be expressed as
\begin{eqnarray}
\mu_{i,V} =\frac{\partial \epsilon_v}{\partial k_i^f}\frac{\partial k_i^f}{\partial n_i}+\frac{\partial \epsilon_v}{\partial M_i}\frac{\partial M_i}{\partial n_i}.
\end{eqnarray}

SQM is assumed to be neutrino-free and composed of $u$, $d$ and $s$ quarks and leptons (electrons and muons) in beta-equilibrium with electric charge neutrality, and the properties of SQM can be obtained under $\beta$-equilibrium condition and electric charge neutrality, i.e.,
\begin{eqnarray}
\mu_d=\mu_s,~~~\mu_d=\mu_u+\mu_e,
\end{eqnarray}
and
\begin{eqnarray}
\frac{2}{3}n_u-\frac{1}{3}n_d-\frac{1}{3}n_s -n_e-n_{\mu} =0.
\end{eqnarray}
For the leptons, we use $\mu_l=\sqrt{(k_l^f)^2+m_l^2}$ to calculate the chemical potential, where $k_l^f=(3\pi^2n_l)^{\frac{1}{3}}$ is the Fermion momentum for leptons ($e$ and $\mu$). And the total pressure for SQM can be obtained from the thermodynamic relation, i.e.,
\begin{eqnarray}
P=&-\epsilon+\sum\limits_{j=u,d,s,l} n_j \mu_j,
\end{eqnarray}
where $\epsilon=\epsilon_k+\epsilon_v$ is the total energy density for SQM.

\section{Results and discussions}\label{Results}
\subsection{Quark matter symmetry energy and isospin splitting of the quark mass}

\begin{figure}[tbp]
\includegraphics[scale=0.32]{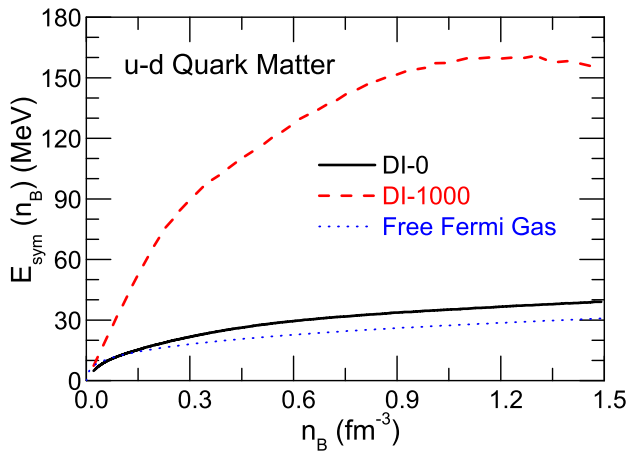}
\caption{(Color online) The symmetry energy of the two flavor $u$-$d$ quark matter as a function of
baryon number density in the ICQM model with DI-0 and DI-1000. The result of a free quark gas is also plotted for comparison.}
\label{Esym}
\end{figure}

Shown in Fig.~\ref{Esym} is the density dependence of the symmetry energy for the two-flavor $u$-$d$ quark matter in the ICQM model with {$D_I=0$ and $\nu_{ud}=\nu_s=0.53$ (denoted as DI-0, see Section IV C for the details) and {$D_I=1000~\text{MeV}~\text{fm}^{3\alpha}$} and $\nu_{ud}=\nu_s=0.68$ (denoted as DI-1000, see Section IV C for the details)}. For comparison, the corresponding result of a free quark gas is also included. {Here, we mainly focus on the case of $\nu_{ud} = \nu_s = \nu$ for simplicity.}
One can see from Fig.~\ref{Esym} that for DI-0 (i.e., the original CQM model), the quark matter symmetry energy at baryon density of $1.5$ fm$^{-3}$ (around $10n_0$) reaches $39$ MeV, while the value is $155$ MeV with DI-1000 and $31$ MeV for the free quark gas. Furthermore, the value of the symmetry energy at nuclear matter normal density $0.17$ fm$^{-3}$ is $59$ MeV for DI-1000, $16$ MeV for DI-0, and $15$ MeV for the free quark gas. These results indicate that DI-1000 leads to a larger value of quark matter symmetry energy (about two times nuclear matter symmetry energy around normal nuclear density) due to a stronger isospin dependent scalar interaction with a large $D_I$.

In addition, the parameters $\alpha$ and $\beta$ control the shape of the density dependence of the quark matter symmetry energy, and indeed the choice of $\alpha=1.5$ and $\beta=1~\text{fm}^3$ in the ICQM model leads to a quark matter symmetry energy having a similar density dependence of the symmetry energy from the free quark gas or the conventional NJL model~\citep{Chu14}. {We can also obtain the values of the density slope parameter of the symmetry energy $L=3n_0 \frac{d E_{sym}(n_B)}{d n_B}|_{n_B=n_0}$ at nuclear matter saturation density $n_0$ as $24.33$ MeV for DI-0, $156.32$ MeV for DI-1000, and $14.93$ MeV for the free quark gas.  }

\begin{figure}[tbp]
\includegraphics[scale=0.45]{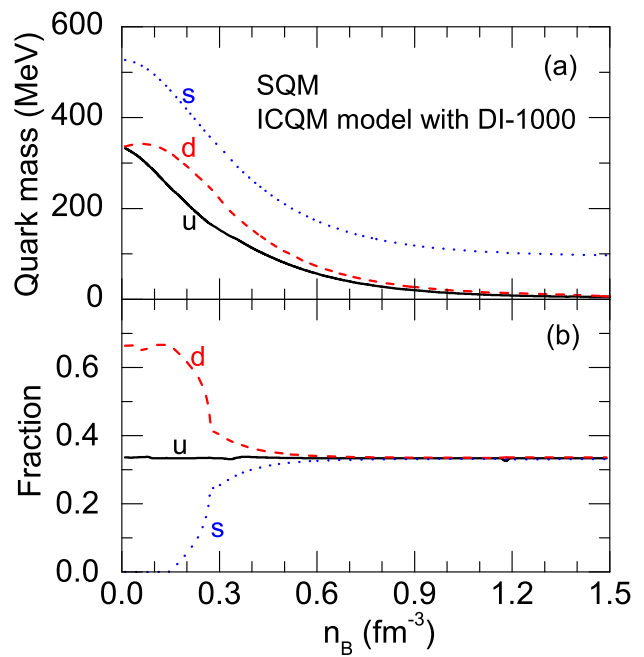}
\caption{(Color online) The quark mass (a) and fraction (b) as a function of the baryon
density in SQM within the ICQM model with DI-1000.}
\label{fraction}
\end{figure}

Shown in Fig.~\ref{fraction}(a) is the density dependence of the quark mass in SQM within the ICQM model with DI-1000. It can be seen that the quark mass decreases drastically as the baryon density increases and the $u$ and $d$ quarks smoothly restore chiral symmetry at high densities. One also sees that there is a clear isospin splitting in the $u$ and $d$ quark masses in SQM, with a heavier $d$ quark mass while a lighter $u$ quark mass. In addition, it is seen that the isospin splitting is strong at lower densities but becomes weaker and weaker and disappears at higher densities. This is due to the fact that the isospin asymmetry in SQM becomes weaker and weaker and disappears at higher densities, as shown in Fig.~\ref{fraction}(b) where the quark fraction is plotted as a function of the baryon density in SQM within the ICQM model with DI-1000. One can find that the $d$ quark fraction is higher than the fractions of $u$ and $s$ quarks at lower densities, then the $d$ quark fraction decreases while the $s$ quark fraction increases as the baryon density increases, and at last the $u$, $d$ and $s$ quark fractions become essentially equal and approach to about $0.33$ when $n_B>0.6$ fm$^{-3}$. This feature is consistent with the picture of the color-flavor-lock (CFL) state~\cite{Raj01}.

\begin{figure}[tbp]
\includegraphics[scale=0.32]{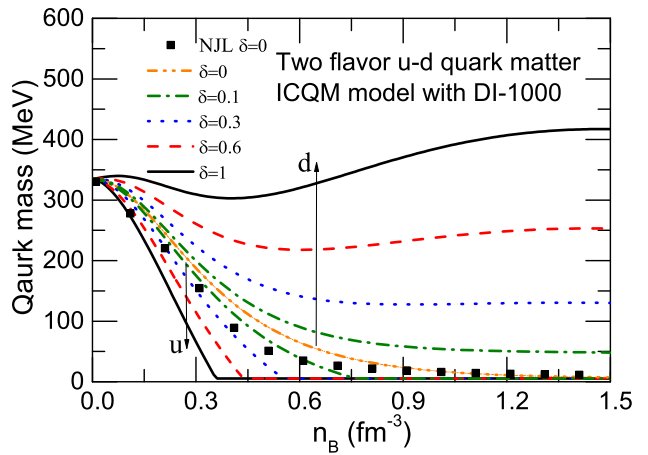}
\caption{(Color online) The quark mass as a function of the baryon
density in the two flavor $u$-$d$ quark matter of $\delta=0$, $0.1$, $0.3$, $0.6$ and $1.0$ within the ICQM model with DI-1000.}
\label{miso}
\end{figure}

To further investigate the isospin splitting of quark mass in asymmetric quark matter, we calculate the quark mass as a function of the baryon density at different values of isospin asymmetry, i.e., $\delta=0$, $0.1$, $0.3$, $0.6$ and $1.0$, in two-flavor $u$-$d$ quark matter within the ICQM model with DI-1000, and the results are shown in Fig.~\ref{miso}. For comparison, we also include in Fig.~\ref{miso} the results for $\delta=0$ from the conventional NJL model~\cite{Hatsuda94}.
It is interesting to see that in the $d$-rich two flavor $u$-$d$ quark matter (i.e., $\delta >0$), the $d$ quark is generally heavier than the $u$ quark, and the isospin splitting increases with the increment of isospin asymmetry $\delta$, namely, the $u$ ($d$) quark mass at a fixed density decreases (increases) as $\delta$ increases. In particular, while the $u$ quarks restore chiral symmetry at a certain density value depending on $\delta$, the $d$ quarks seem to stabilize its mass at high density and cannot restore chiral symmetry at least up to the density $1.5$ fm$^{-3}$ considered here. This feature about the isospin dependence of the chiral symmetry restoration is consistent with the conclusion on pure neutron matter calculations based on QCD sum rules~\cite{Cai17}. Our results will have important implications on the QCD phase diagram at finite baryon density under extreme isospin condition~\cite{Liu16}.

\subsection{Stability of SQM}

Following Farhi and Jaffe~\citep{Farhi84}, the absolute stability requires that the minimum energy per baryon of SQM should be less than the minimum energy per baryon of the observed stable nuclei, which is $930$ MeV, and the minimum energy per baryon of the $\beta$-equilibrium $u$-$d$ quark matter should be larger than $930$ MeV to be consistent with the standard nuclear physics. The stability conditions usually put very strong constraints on the value of the parameters in phenomenological quark matter models.

\begin{figure}[tbp]
\includegraphics[scale=0.32]{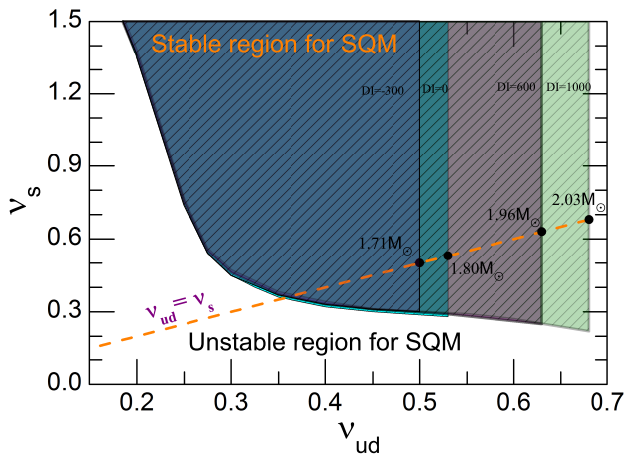}
\caption{(Color online) Stability region in the $\nu_{ud}$-$\nu_{s}$ plane for SQM in the ICQM model with $D_I=-300~\text{MeV}~\text{fm}^{3\alpha}$, $D_I=0$ (the case of the original CQM model), $D_I=600~\text{MeV}~\text{fm}^{3\alpha}$ and $D_I=1000~\text{MeV}~\text{fm}^{3\alpha}$. The maximum mass of QS for each $D_I$ with $\nu_{ud}=\nu_{s}$ is indicated.}
\label{stability}
\end{figure}

Figure~\ref{stability} shows the stability region in the $\nu_{ud}$-$\nu_{s}$ plane in which the absolute stability for SQM is considered for the ICQM model. We choose different vales of the isospin parameters $D_I$, namely, $D_I=-300~\text{MeV}~\text{fm}^{3\alpha}$, $D_I=0$ (the case of the original CQM model), $D_I=600~\text{MeV}~\text{fm}^{3\alpha}$ and $D_I=1000~\text{MeV}~\text{fm}^{3\alpha}$ to consider the isospin effects on the stability region for SQM. After detailed calculations, we find that the minimum energy per baryon of the $\beta$-equilibrium $u$-$d$ quark matter decreases with the increment of $\nu_{ud}$, and then the maximum value of $\nu_{ud}$ for $D_I=-300~\text{MeV}~\text{fm}^{3\alpha}$, $D_I=0$, $D_I=600~\text{MeV}~\text{fm}^{3\alpha}$ and $D_I=1000~\text{MeV}~\text{fm}^{3\alpha}$ under the absolutely stable condition (when the minimum energy per baryon for the $\beta$-equilibrium $u$-$d$ quark matter is exactly $930$ MeV, and meanwhile the minimum energy per baryon for SQM is less than $930$ MeV) is $0.50$, $0.53$, $0.63$ and $0.68$, respectively. Furthermore, we also find that the minimum energy per baryon of SQM increases when $\nu_s$ decreases for a fixed $\nu_{ud}$, and then we can obtain the lower limit boundary of the stability region for SQM (when the minimum energy per baryon is exactly $930$ MeV for SQM, and meanwhile the minimum energy per baryon for the $\beta$-equilibrium $u$-$d$ quark matter is larger than $930$ MeV) for $D_I=-300~\text{MeV}~\text{fm}^{3\alpha}$, $D_I=0$, $D_I=600~\text{MeV}~\text{fm}^{3\alpha}$ and $D_I=1000~\text{MeV}~\text{fm}^{3\alpha}$, as shown in Fig.~\ref{stability}. It is seen that the right boundary for this stability region moves toward the right direction (i.e., larger $\nu_{ud}$) when the $D_I$ increases, while the left boundary and the lower boundary for these four different $D_I$ cases are almost the same. These features imply that increasing the $D_I$ value generally leads to a larger stability region in the $\nu_{ud}$-$\nu_{s}$ plane.

\begin{figure}[tbp]
\includegraphics[scale=0.34]{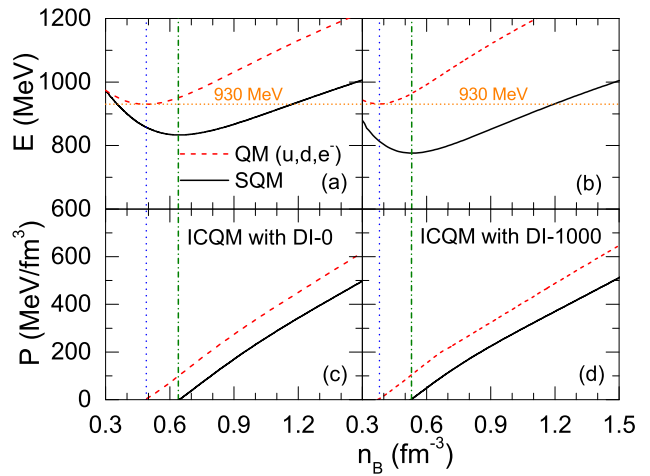}
\caption{(Color online) Energy per baryon and the corresponding pressure
as functions of the baryon density for SQM and two-flavor $u$-$d$
quark matter in $\beta$-equilibrium within the ICQM model with DI-0 and DI-1000.}
\label{EOS}
\end{figure}

To more clearly see the thermodynamic self-consistency and the absolute stability of SQM in our present model, we show in Fig.~\ref{EOS} the energy per baryon and the corresponding pressure as functions of the baryon number density for SQM and $u$-$d$ quark matter in $\beta$-equilibrium condition within the ICQM model with DI-0 and DI-1000.  As one can see from all the cases in Fig.~\ref{EOS}, the minimum energy per baryon of the beta-equilibrium two-flavor $u$-$d$ quark matter is larger than $930$ MeV, while the corresponding values of SQM is less than $930$ MeV, satisfying the requirement of the absolute stability. One can also see in Fig.~\ref{EOS} that the minimum energy per baryon is exactly the zero-pressure density for both the parameter sets DI-0 and DI-1000, consistent with the requirement of thermodynamical self-consistency.

\subsection{Quark stars}

Based on the absolute stability region for SQM, we can then calculate the maximum mass of QS with the parameter sets under absolutely stable condition, and our results indicate: a) The maximum mass of QS increases with the increment of $\nu_s$, when $\nu_{ud}$ is fixed for all $D_I$ cases; and b) The maximum mass of QS increases with the increment of $\nu_{ud}$, when $\nu_{s}$ is fixed for all $D_I$ cases.
Since we mainly focus on the case of $\nu_{ud} = \nu_s = \nu$ in the present work, we plot the line with $\nu_{ud}=\nu_s$ in Fig.~\ref{stability} and indicate the maximum mass of QS for each considered $D_I$. One can see that the maximum mass for quark star along the $\nu_{ud}=\nu_s$ line increases with $\nu_{ud}$ ($D_I$). One can also see that the maximum mass of QS with the parameter sets along this line will not reach the mass of $2.01\pm0.04~M_{\odot}$ until $D_I=1000~\text{MeV}~\text{fm}^{3\alpha}$ and $\nu_{ud}=\nu_s=0.68$, which gives the lower limit of $D_I$ for the $\nu_{ud}=\nu_s$ case to describe a $2~M_{\odot}$ QS within the ICQM model. In addition, one can also see that the maximum mass of QS for $D_I=0$ with $\nu_{ud}=\nu_s=0.53$ is only $1.80~M_{\odot}$, which is the heaviest QS under absolutely stable condition within the CQM model~\citep{Dey98}. In the following, for convenience, we denote the parameter set with $D_I=1000~\text{MeV}~\text{fm}^{3\alpha}$ and $\nu_{ud}=\nu_s=0.68$ as DI-1000, while the one with $D_I=0~\text{MeV}$ and $\nu_{ud}=\nu_s=0.53$ as DI-0. {Similarly, we denote the parameter set with $D_I=-300~\text{MeV}~\text{fm}^{3\alpha}$ and $\nu_{ud}=\nu_s=0.50$ as DI-m300, and that with $D_I=600~\text{MeV}~\text{fm}^{3\alpha}$ and $\nu_{ud}=\nu_s=0.63$ as DI-600.}

\begin{figure}[tbp]
\includegraphics[scale=0.32]{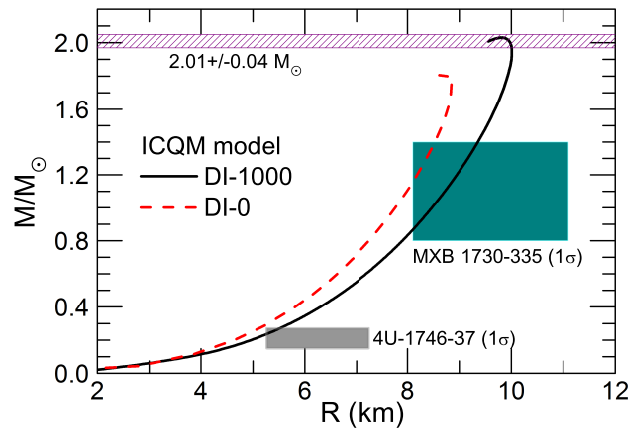}
\caption{(Color online) Mass-Radius relation within the ICQM model with DI-0 (corresponding to the original CQM model) and DI-1000. For comparison, the mass of $2.01\pm0.04~M_{\odot}$ for PSR J0348+0432~\citep{Antoniadis13}, the $M=1.1\pm0.3~M_{\odot}$ and $R=9.6\pm1.5$ km($1\sigma$)~\citep{Sala12} for MXB 1730-335 (dark cyan box), and $M=0.21\pm0.06~M_{\odot}$ and $R=6.26\pm0.99$ km~\citep{Li15} for 4U 1746-37 (grey box), are also included.}
\label{MR}
\end{figure}

Shown in Fig.~\ref{MR} is the mass-radius relation for static QSs within the ICQM model with DI-0 and DI-1000 by solving the Tolman-Oppenheimer-Volkov equation~\citep{Open39}. The pulsar PSR J0348+0432 with a mass of $2.01\pm0.04~M_{\odot}$~\citep{Antoniadis13}, which sets a new record for the maximum mass of pulsars, is indicated in Fig.~\ref{MR}.
In addition, the shadowed box with dark cyan color in Fig.~\ref{MR} is the recently measured mass and radius of the pulsars in the rapid burster MXB 1730-335, which gives the region constrained to be $M=1.1\pm0.3~M_{\odot}$ and $R=9.6\pm1.5$ km ($1\sigma$) by the analysis of Swift/XRT time-resolved spectra of the burst~\citep{Sala12}, while the shadowed box with grey color in Fig.~\ref{MR} is the $1\sigma$ confidence level of mass and radius of object 4U 1746-37 with the corresponding range being $M=0.21\pm0.06~M_{\odot}$ and $R=6.26\pm0.99$ km~\citep{Li15}. We can find that the results of the ICQM model with DI-1000 is consistent with the observations, while the results from the original CQM model (DI-0) cannot describe the pulsar PSR J0348+0432 and 4U 1746-37. {On the other hand, we would like to point out that the accurate determination of the radius for pulsars is highly nontrivial and still remains a big challenge (see, e.g., Refs.~\cite{Lattimer13,Lattimer14,Guillot13,Guillot14,Menezes17,Hebeler10,Guver13}). Therefore, our present results suggest that in the ICQM model, the symmetry energy of the two-flavor $u$-$d$ quark matter should be at least about four times stronger than that of a free quark gas to describe the heavy pulsar PSR J0348+0432 as a quark star. This also suggests that the quark mass should exhibit a stronger isospin splitting in isospin asymmetric quark matter.}.

\subsection{Effects of density dependence of the quark mass}

As mentioned earlier, $\nu_{u}$, $\nu_{d}$ and $\nu_s$ in Eq.~(\ref{mqiso}) are phenomenological parameters introduced to control the density dependence of the quark mass in the ICQM model, and their values are still unclear. In the above calculations, for simplicity, we have assumed  $\nu_{u} = \nu_{d} = \nu_s = \nu$, just like in the original CQM model. In the ICQM model, the term with parameter $D_I$ in Eq.~(\ref{mqiso}) is introduced to consider the isospin dependence of the $u$ and $d$ quark masses and at the same time to keep the isospin symmetry of QCD, which automatically leads to different density dependence of the $u$ and $d$ quark masses in isospin asymmetric quark matter.
For more general case, the $\nu_{ud}$ and $\nu_s$ could be different, which can  be seen in the SU(3) NJL model.

As we pointed out before, the maximum mass of QS depends on the values of $\nu_{ud}$ and $\nu_s$. If the $\nu_{ud}$ and $\nu_s$ are allowed to have different values, then the parameter $D_I$ may have different values to describe the heavy pulsar PSR J0348+0432 with a mass of $2.01\pm0.04~M_{\odot}$ as a quark star. For example, we can obtain $2M_{\odot}$ QSs by using $D_I=-300~\text{MeV}~\text{fm}^{3\alpha}$ with $\nu_{ud}=0.5$ and $\nu_s=1.2$ (denoted as DI$^{*}$-m300), or $D_I=0$ with $\nu_{ud}=0.53$ and $\nu_s=1$ (denoted as DI$^{*}$-0), or $D_I=600~\text{MeV}~\text{fm}^{3\alpha}$ with $\nu_{ud}=0.63$ and $\nu_s=0.7$ (denoted as DI$^{*}$-600), where $\nu_s$ is the lower-limit value under absolutely stable condition to support $2M_{\odot}$ QSs in each case. The maximum mass and the corresponding radius of QSs with these parameter sets and DI-1000 are summarized Table~\ref{MRT}. The maximum mass of QS can be further enhanced as the $\nu_s$ increases.

\begin{table}
  \centering
  \caption{The maximum mass and the corresponding radius of QSs in the ICQM model with DI$^{*}$-0, DI$^{*}$-m300, DI$^{*}$-600 and DI-1000.}
  \begin{tabular}
  {rccccc}
  \hline
  \hline
  &DI$^{*}$-0 &DI$^{*}$-m300 &DI$^{*}$-600 &DI-1000 \\
  \hline
  $M$ ($M_{\odot}$)&2.04~&2.03~&2.02~&2.03~\\
  \hline
  $R$ (km)&9.795~&9.755~&9.690~&9.815~\\
  \hline
  \end{tabular}
  \label{MRT}
\end{table}

We also calculate the quark matter symmetry energy in two flavor $u$-$d$ quark matter using DI$^{*}$-m300 and DI$^{*}$-600, and we find that the quark matter symmetry energy at $n_B = 1.5$ fm$^{-3}$ is $52$ MeV for DI$^{*}$-m300 and $82$ MeV for DI$^{*}$-600. It should be mentioned that DI$^{*}$-600 predicts a quark matter symmetry energy which is about twice that of a free quark gas or normal quark matter within the conventional NJL model, consistent with the CIDDM model predictions~\cite{Chu14}. In addition, we calculate the values of the magnitude $E_{\text{sym}}(n_0)$ and density slope parameter $L$ of the quark matter symmetry energy at $n_0$ for different sets of parameters, and the results are shown in Table~\ref{EsymL}. It should be noted that for two flavor $u$-$d$ quark matter, DI-m300 gives the same quark matter symmetry energy as DI$^{*}$-m300 since they have the same $\nu_{ud}$. This is also true for DI-0 and DI$^{*}$-0 as well as DI-600 and DI$^{*}$-600. One can find that both $E_{\text{sym}}(n_0)$ and $L$ increases as the magnitude of the isospin parameter $D_I$ becomes larger. It is interesting to see the values of the magnitude $E_{\text{sym}}(n_0)$ and density slope parameter $L$ of the quark matter symmetry energy at $n_0$ with DI$^{*}$-m300 are in very good agreement with the empirical values for nuclear matter symmetry energy (see, e.g., Ref.~\cite{Chen17}). These results indicate that varying $\nu_{ud}$ and $\nu_s$ can significantly change the thermodynamic properties of SQM and asymmetric quark matter in the ICQM model.

\begin{table}
  \centering
    \caption{The magnitude $E_{\text{sym}}(n_0)$ and the density slope parameter $L$ of the quark matter symmetry energy in two flavor $u$-$d$ quark matter at nuclear saturation density $n_0$ in the ICQM model with DI$^{*}$-0, DI$^{*}$-m300, DI$^{*}$-600 and DI-1000. The corresponding results for free Fermi gas (FG) of two flavor $u$-$d$ quark matter are also included for comparison.}
  \begin{tabular}
  {rccccc}
  \hline
  \hline
  &FG &DI$^{*}$-0 &DI$^{*}$-m300 &DI$^{*}$-600 &DI-1000\\
  \hline
  $L$ (MeV)&14.93~&24.33~&63.26~&109.23~&156.32~\\
  \hline
    $E_{\text{sym}}(n_0)$ (MeV)&14.91~&16.41~&29.18~&42.10~&59.20~\\
  \hline
  \end{tabular}
\label{EsymL}
\end{table}

\section{Summary and Conclusions}

We have extended the CQM model, in which the quark confinement and asymptotic freedom are described by using a vector potential (Richardson potential) and the chiral symmetry is modeled by a density dependent quark mass, to include isospin dependence of the quark mass. Within the extended ICQM model, we have investigated the absolutely stable region of SQM in beta-equilibrium condition, the mass-radius relation for QSs, the quark matter symmetry energy, and the isospin splitting of quark mass. We have also discussed the effects of the density dependence of the quark mass on our results.
We have found
although the recently discovered heavy pulsars PSR J1614.2230 and PSR J0348+0432 with
masses around $2M_{\odot}$ cannot be quark stars within the original CQM model, they can be
well described by quark stars in the ICQM model by considering isospin and density dependence of the quark mass.

If we assume $u$, $d$ and $s$ quarks have the same density dependence for the quark mass in isospin symmetric quark matter, we have found that, in order to describe the heavy pulsars PSR J1614.2230 and PSR J0348+0432 as quark stars, the isospin dependence of quark mass should be strong enough so that the quark matter symmetry energy is at leas about four times that of a free quark gas. In this case, our results suggest a strong isospin splitting of the quark mass with the $u$ ($d$) quark mass at a fixed density decreasing (increasing) as isospin asymmetry $\delta$ increases. In particular, we have found that while the $u$ quarks restore chiral symmetry at a certain density value depending on isospin asymmetry $\delta$, the $d$ quarks seem to stabilize its mass at high density and cannot restore chiral symmetry at least up to the density $1.5$ fm$^{-3}$. Our results will thus have important implications on the isospin dependence of the QCD phase diagram at finite baryon density.

In addition, we have found that the density dependence of $u$, $d$ and $s$ quark masses can significantly influence the properties of SQM and thus the QSs. Our results have demonstrated that the existence of large quark stars with mass around $2M_{\odot}$ can put important constraints on isospin and density dependence of the quark mass as well as the quark matter symmetry energy.

\section*{Acknowledgments}
We thank Gang Guo and Zhen Zhang for helpful discussions. This work was supported in part by the Major State Basic Research Development Program (973 Program) in China under
Contract Nos. 2015CB856904 and 2013CB834405, the National Natural Science Foundation of China under Grant
Nos. 11625521, 11275125, 11135011 and 11505100, the Program for Professor of Special Appointment (Eastern Scholar) at Shanghai Institutions of Higher Learning, Key Laboratory for Particle Physics, Astrophysics and Cosmology, Ministry of Education, China, the Science and Technology Commission of Shanghai Municipality (11DZ2260700), and the Shandong Provincial Natural Science Foundation, China (ZR2015AQ007).

\end{document}